\documentclass[prb,aps,twocolumn,superscriptaddress]{revtex4-1}

\usepackage{graphics, bm, psfrag, amsmath, amssymb, epsfig, grffile, float, bbold}
\usepackage{subfigure}
\usepackage{dsfont}
\usepackage{color}
\usepackage{hyperref}
\usepackage{amsfonts,amsthm,mathtools}
\usepackage{latexsym,graphicx}
\usepackage{amscd}
\usepackage{dblfloatfix}
\usepackage{rotating}
\usepackage[percent]{overpic}

\newlength{\intwidth}
\DeclareRobustCommand{\fpint}[2]
   {\mathop{%
      \text{%
        \settowidth{\intwidth}{$\int$}%
        \makebox[0pt][l]{\makebox[\intwidth]{$-$}}%
        $\int_{#1}^{#2}$}}}

\newcommand{\Om}{\Omega} 
\newcommand{\sign}{\mathrm{sgn}} 
\newcommand{\xxa}{: \!}
\newcommand{\xxe}{\!:}

\newcommand{\huz}{\rho_0}
\newcommand{\hvz}{\sigma_0}
\newcommand{\hu}{\rho}
\newcommand{\hv}{\sigma}

\newcommand{\cH}{\mathcal{H}}

\newcommand{\vx}{\boldsymbol{x}}

\newcommand{\Vc}{V_{\mathrm{conf}}} 

\newcommand{\ee}{{\rm e}}
\newcommand{\ii}{{\rm i}}
\newcommand{\dd}{d}

\newcommand{\R}{{\mathbb R}}
\newcommand{\C}{{\mathbb C}}
\newcommand{\Z}{{\mathbb Z}}

\newcommand{\vu}{\boldsymbol{u}}
\newcommand{\vv}{\boldsymbol{v}}

\newcommand{\cT}{\mathcal{T}} 

\renewcommand{\Re}{\mathrm{Re}\hspace{0.09em}}
\renewcommand{\Im}{\mathrm{Im}\hspace{0.09em}}

\newcommand{\nd}{{\phantom\dag}}

\begin{document}

\title{Nonchiral Intermediate long-wave equation and interedge effects in narrow quantum Hall systems}

\author{Bjorn K. Berntson}
\email{bbernts@kth.se}
\affiliation{Department of Mathematics, KTH Royal Institute of Technology, SE-100 44 Stockholm, Sweden}
\author{Edwin Langmann}
\email{langmann@kth.se}
\affiliation{Department of Physics, KTH Royal Institute of Technology, SE-106 91 Stockholm, Sweden}
\author{Jonatan Lenells}
\email{jlenells@kth.se}
\affiliation{Department of Mathematics, KTH Royal Institute of Technology, SE-100 44 Stockholm, Sweden}

\date{\today}

\begin{abstract}
We present a nonchiral version of the Intermediate long-wave (ILW) equation that can model nonlinear waves propagating on two opposite edges of a quantum Hall system, taking into account interedge interactions. 
We obtain exact soliton solutions  governed by the hyperbolic Calogero-Moser-Sutherland (CMS) model, and we give a Lax pair, a Hirota form, and conservation laws for this new equation. 
We also present a periodic nonchiral ILW equation, together with its soliton solutions governed by the elliptic CMS model.
\end{abstract}

\maketitle

\section{Introduction}
One important feature of the Fractional Quantum Hall Effect (FQHE) is the strikingly high accuracy by which the Hall conductance, $\sigma_{\mathrm H}$,  is measured in units of the inverse von Klitzing constant, $e^2/h$.\cite{ES90}  
Therefore, satisfactory explanations of these FQHE measurements, $\sigma_{\mathrm H} h/e^2=\frac13,\frac25,\frac37,\ldots$, must be based on exact analytic arguments, and theories of the FQHE have close connections to integrable systems.  
Two important classes of integrable systems which are seemingly very different but which are both connected with the FQHE are (i) Calogero-Moser-Sutherland (CMS)\cite{rem1} models describing FQHE edge states,\cite{AI94,I95,AMOS95,ES98,CL99,BH08,AL17} and (ii) soliton equations of Benjamin-Ono (BO) type describing the dynamics of nonlinear waves propagating along FQHE edges\cite{AW05,BAW06,W12} (background on the soliton equations appearing in this paper can be found in Section~\ref{sec:applB}). 
These systems are related by a fundamental correspondence between CMS systems and BO-type soliton equations, which provides the basis for a mathematically precise derivation of hydrodynamic descriptions of CMS systems.\cite{P95,SG08,SAX08,ABW09}
It is worth noting that this subject has recently received considerable attention in the context of nonequilibrium physics.\cite{CDY16,BCDF16,D19,S19}

While the CMS-BO correspondence has been successfully used to understand FQHE physics, it is incomplete. 
Indeed, CMS systems come in four types: (I) rational, (II) trigonometric, (III) hyperbolic, and (IV) elliptic,\cite{OP81,OP83}  and while the soliton equations related to the rational and trigonometric cases are well-understood  since a long time, \cite{SG08,SAX08,ABW09} soliton equations related to the hyperbolic and elliptic cases were only recently identified as the Intermediate long-wave (ILW) equation and the periodic ILW equation, respectively.\cite{BSTV14,ZZ18,GKKV19}
However, as we will show in this paper, the latter two soliton equation are not unique: there are other equations which are more interesting in that they are of a different kind and describe new physics.

The correspondence between CMS and BO systems exists both at the classical\cite{SAX08,ABW09} and at the quantum\cite{AW05,SG08} level, and we consider both.  
As will be explained, we discovered the quantum elliptic version of the soliton equation presented in this paper from a second quantization of the quantum elliptic CMS model.\cite{L00,L04} 
However, the exact results on the solution of this equation presented in this paper are restricted to the classical case for simplicity.  
We first give and prove our results in the hyperbolic case; the generalization to the elliptic case is surprisingly easy, as will be shown later on. 

\paragraph*{Plan.} In Section~\ref{sec:heuristic}, we give a heuristic argument motivating a generalization of the BO equation that describes coupled nonlinear waves propagating in opposite directions, 
 and we present this so-called nonchiral ILW equation in Section~\ref{sec:ILW}. Our quantum results can be found in Section~\ref{sec:Quantum}: First, the relation between the quantum elliptic CMS model and a quantum version of the nonchiral ILW equation is presented (Section~\ref{sec:CFTA}); second, a detailed motivation of our proposal that the quantum nonchiral ILW equation can describe nonlinear waves propagating on two opposite edges of a FQHE system boundary and taking into account interactions between different edges is given, including a review of the relevant background (Section~\ref{sec:CFTB}). Results that prove that the classical nonchiral ILW equations are exactly solvable can be found in Sections~\ref{sec:hyp} (hyperbolic case) and \ref{sec:ell} (elliptic case). 
In Section~\ref{sec:applications}, we shortly recall the application of the BO- and ILW equations to nonlinear water waves, and we present a simple physical argument suggesting that the nonchiral ILW equation is also relevant in that context.
We conclude with final remarks in Section~\ref{sec:final}. Appendix~\ref{app:A} provides mathematical details, and Appendix~\ref{app:B} shortly explains numerical computations we performed to test our exact analytic solutions. 

\section{Classical physics description}
\label{sec:Classical}
As explained in the Section~\ref{sec:Quantum}, we discovered the quantum version of the nonchiral ILW in the context of the FQHE. 
However, for simplicity, we first present in this section a simpler heuristic argument on the classical level which leads to the classical version of this equation. 
As elaborated in Section~\ref{sec:applications} in one example, this heuristic argument can be straightforwardly adapted to other situations, 
suggesting that the nonchiral ILW will also find other applications in physics.

\subsection{Heuristic motivation}
\label{sec:heuristic}
\begin{figure}[!htbp]
\centering
\vspace{.2cm}
\begin{overpic}[width=.3\textwidth]{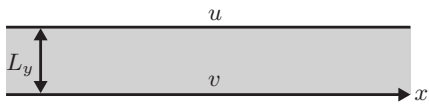}
      \put(50,4.5){\small $v$} 
      \put(50,20.8){\small $u$}  
      \put(101,1){\small $x$} 
      \put(0,8.5){\small $L_y$} 
     \end{overpic}
     \vspace{-.2cm}
\caption{Schematic picture of a narrow FQHE system with two edges carrying the two nonlinear waves $u(x,t)$ and $v(x,t)$.}
\label{Fig1} 
\end{figure}

The CMS models can be defined by Newton's equations
\begin{equation} 
\label{Newton} 
\ddot z_j = -\sum_{k\neq j}^N 4V'(z_j-z_k)\quad (j=1,\ldots,N) 
\end{equation} 
where the two-body interaction potential is $V(r)=r^{-2}$ in the rational case and $V(r) = (\pi/L)^2 \sin^{-2}(\pi r/L)$, $L>0$,  in the trigonometric case\cite{OP81} (the arguments in this paragraph apply to both cases).\cite{remg} 
Eq.\ \eqref{Newton} describes an arbitrary number, $N$, of interacting particles with positions $z_j\equiv z_j(t)$ at time $t$.  
While one often restricts to real positions when interpreting the CMS model as a dynamical system, one has to allow for complex $z_j$ when studying the relation to the BO equation;\cite{SAX08,ABW09} this generalization preserves the integrability.\cite{C01}  
The CMS model is invariant under the parity transformation $P: z_j\to -z_j$ for all $j$. 
However, the corresponding BO equation is not parity-invariant: it is given by $u_t + 2 u u_x + Hu_{xx}=0$, where $u\equiv u(x,t)$ and $H$ is the Hilbert transform (in the rational case, $(Hf)(x)=(1/\pi)\fpint{\R}{}(x'-x)^{-1}f(x')\dd{x'}$, where $\fpint{}{}$ denotes the usual Cauchy principle value integral), and under the parity transformation $P: u(x,t)\to u(-x,t)\equiv v(x,t)$,  it changes to $v_t-2v v_x - Hv_{xx}=0$. 
This mismatch of symmetry is paradoxical at first sight, but the paradox is resolved by interpreting $u$ as a wave propagating on one edge of a FQHE system and noting that, in general, there is another edge far away carrying another wave $v$. 
Thus, actually, the rational CMS model corresponds to two uncoupled BO equations for $u$ and $v$. 
This   system of equations is invariant under a parity transformation interchanging $u$ and $v$, 
\begin{equation} 
\label{P} 
P: [u(x,t), v(x,t)] \to [v(-x,t),u(-x,t)].
\end{equation} 
It is peculiar that these two BO equations are uncoupled, and it is for this reason that one can reduce the system to a single equation, ignoring the other. 
While this uncoupling is reasonable if the two edges are infinitely far apart, it is natural to ask what would happen  
if the two edges are parallel and close together; see~Fig.~\ref{Fig1}.  
In this case, one would expect that the nonlinear waves propagating on the two edges interact. 
We now give a simple heuristic argument to suggest that the hyperbolic CMS model can describe this situation. 

\vspace{-5pt}

The hyperbolic CMS model can be defined by Newton's equations \eqref{Newton} with the interaction potential 
\begin{equation} 
\label{V}
V(r) = \sum_{n\in\Z}\frac1{(r+2\ii\delta n)^2}=\left(\frac{\pi}{2\delta}\right)^2\sinh^{-2}\left(\frac{\pi }{2\delta} r\right), 
\end{equation} 
where $\delta>0$ is an arbitrary length parameter. 
Dividing the particle positions $z_j$ into two groups and shifting the ones in the second group by the imaginary half-period, $w_k\equiv z_{k-N_1}+\ii \delta$ for $k=1,\ldots,N_2\equiv N-N_1$, with $1<N_1<N$, we can write these Newton's equations as  
\begin{equation} 
\label{Newtonzw}
\begin{split} 
&\ddot z_j = -\sum_{j'\neq j}^{N_1} 4V'(z_j-z_{j'}) - \sum_{k=1}^{N_2}4\tilde{V}'(z_j-w_k) ,\\
&\ddot w_k = -\sum_{k'\neq k}^{N_2} 4V'(w_k-w_{k'}) - \sum_{j=1}^{N_1}4\tilde{V}'(w_k-z_j) 
\end{split} 
\end{equation} 
for $j=1,\ldots,N_1$ and $k=1,\ldots,N_2$, with 
\begin{equation} 
\tilde V(r)\equiv V(r-\ii\delta)=-\left(\frac{\pi}{2\delta}\right)^2\cosh^{-2}\left(\frac{\pi }{2\delta} r\right). 
\end{equation} 
This can be interpreted as a model of two kinds of particles, $z_j$ and $w_k$, in which particles of the same kind interact via the singular repulsive two-body potential $V$, whereas particles of different kinds interact via the weakly attractive nonsingular potential $\tilde{V}$. 
We interpret $\delta$ as a parameter of the same order of magnitude as the distance $L_y$ between the two edges of the FQHE system; see Fig.~\ref{Fig1}. In the rational limit $\delta\to\infty$, we have $\tilde{V}\to 0$, so particles of different types do not interact and the two corresponding soliton equations for $u$ and $v$ decouple;  for finite $\delta$, the system is coupled.

\subsection{Nonchiral ILW equation.}
\label{sec:ILW} 
In the hyperbolic case, the two-component generalization of the BO equation we present in this paper is given by 
\begin{equation} 
\label{2ILW} 
\begin{split} 
&u_t + 2 u u_x + Tu_{xx}+\tilde{T}v_{xx}=0,\\
&v_t - 2 v v_x - Tv_{xx}-\tilde{T}u_{xx}=0
\end{split} 
\end{equation} 
for $u=u(x,t)$ and $v=v(x,t)$, with 
\begin{equation} 
\label{TT} 
\begin{split} 
& (Tf)(x) \equiv \frac{1}{2\delta}\fpint{\R}{} \coth\left(\frac{\pi}{2\delta}(x'-x)\right)f(x')\dd{x}',\\
& (\tilde{T}f)(x) \equiv \frac{1}{2\delta}\int_{\R}{} \tanh\left(\frac{\pi}{2\delta}(x'-x)\right)f(x')\dd{x}'. 
\end{split} 
\end{equation} 
The standard ILW equation is $u_t +2 uu_x + Tu_{xx}=0$;\cite{J77,KSA81,rem3} it reduces to the BO equation in the limit $\delta\to\infty$.
Thus, if one drops the $\tilde{T}$-terms, \eqref{2ILW} corresponds to a system of uncoupled ILW equations generalizing the system of uncoupled BO equations discussed above. 
However, due to the presence of the $\tilde{T}$-terms, the nonlinear waves $u$ and $v$ interact. 
For this reason, and since equation \eqref{2ILW} is invariant under the parity transformation \eqref{P}, we call it the {\em nonchiral ILW equation}; 
another motivation for this name is its relation to a nonchiral conformal field theory explained in Section~\ref{sec:CFTA}.

For later reference, we also give the nonchiral version of the periodic ILW equation:\cite{AFSS82} it is defined by \eqref{2ILW} but with the integral operators  
\begin{equation}
\label{TTe}
\begin{split} 
&(Tf)(x) = \frac1{\pi}\fpint{-L/2}{L/2} \zeta_1(x'-x)f(x')\dd{x}',\\
&(\tilde{T}f)(x)=\frac1{\pi}\int_{-L/2}^{L/2} \zeta_1(x'-x+\ii\delta)f(x')\dd{x}', 
\end{split} 
\end{equation}
where 
\begin{equation} 
\label{MLe} 
\zeta_1(z) = \frac{\pi}{L}\lim_{M\to\infty}\sum_{n=-M}^M \cot\left(\frac{\pi}{L}(z-2\ii n\delta) \right)
\end{equation} 
is equal to the Weierstrass elliptic $\zeta$-function with periods $(L,2\ii\delta)$, up to a term proportional to $z$.\cite{WW40}
To see that the operators in \eqref{TTe} are natural periodic generalizations of the ones in \eqref{TT}, we recall that 
\begin{equation}
\frac{\pi}{2\delta}\coth\left(\frac{\pi }{2\delta}z\right) = \lim_{M\to\infty}\sum_{n=-M}^M \frac1{z- 2\ii\delta n}.
\end{equation}

\section{Quantum physics description}
\label{sec:Quantum}
It is known that the edge excitations in a FQHE system can be described by a conformal field theory (CFT) of chiral bosons,\cite{W90} and that this CFT accommodates a quantum version of the BO equation\cite{AW05,SG08} which, at the same time, provides a second quantization of the trigonometric CMS system.\cite{AI94,I95,AMOS95,ES98,CL99,BH08,AL17} This CFT is a nonlinear, exactly solvable system that can describe universal features of FQHE physics; in particular, as proposed by Wiegmann,\cite{W12} this description implies that {\em the dynamics of FQHE edge states is essentially nonlinear, and it features fractionally charged solitons with charges determined by the filling level, $\nu$}.

In this section, we explain how these results generalize to the elliptic case, and how this led us to the nonchiral ILW equation (Section~\ref{sec:CFTA}). 
We also substantiate our proposal that the (quantum version of the) nonchiral ILW equation can describe the interaction of nonlinear waves on two edges in a FQHE system, taking into account interedge effects (Section~\ref{sec:CFTB}). 
This section can be skipped without loss of continuity. 

\subsection{CFT and nonchiral ILW equation} 
\label{sec:CFTA} 
The (quantum) elliptic CMS system is defined by the Hamiltonian
\begin{equation} 
\label{eCS} 
H_N(\vx)  = -\frac12 \sum_{j=1}^N \frac{\partial^2}{\partial x_j^2} + \sum_{1\leq j< k\leq N}g(g-1) \wp_1(x_j-x_k)
\end{equation} 
where 
\begin{equation} 
\label{wp1} 
\wp_1(x) = \sum_{n\in\Z} \frac{\left(\frac{\pi}{L}\right)^2}{\sin^2\left(\frac{\pi}{L}(x-2\ii n\delta) \right)}
\end{equation} 
equals the Weierstrass elliptic $\wp$-function with periods $(L,2\ii\delta)$, up to an additive constant\cite{WW40} (we use units such that $2m=\hbar=1$). 
The parameter $g>0$ is the coupling constant, and $(g-1)$ is to be interpreted as  $(g-\hbar)$, i.e., $g(g-1)\to g^2$ in the classical limit. 
Thus, for $g=2$, the Hamiltonian in \eqref{eCS} defines the quantum analogue of the classical model defined by Newton's equations in \eqref{Newton} for $V(x)=\wp_1(x)$. 
It is important to note that $g$ is an essential parameter in the quantum case, different from the classical case where we can set $g=2$ without loss of generality.\cite{remg} 

The CFT corresponding to the elliptic CMS system can be defined by two chiral boson operators $\huz(x)$ (right-movers) and $\hvz(x)$ (left-movers) labeled by a coordinate $x\in[-L/2,L/2]$ 
on the circle with circumference $L>0$ and satisfying the the commutator relations 
\begin{equation} 
\label{CCR0} 
\begin{split} 
[\huz(x),\huz(x')] &= -2\pi \ii \nu\partial_x\delta(x-x') , \\
[\hvz(x),\hvz(x')] &= 2\pi \ii \nu\partial_x\delta(x-x') , 
\end{split} 
\end{equation} 
and $[\huz(x),\hvz(x')]=0$, with $\nu$ the filling factor of the FQHE system;\cite{W90} 
the latter can be identified with the inverse of the coupling parameter in the corresponding CMS Hamiltonian: $\nu=1/g$.\cite{L00,L04}  
For simplicity, we restrict our discussion to FQHE states where $g=3,5,\ldots$, even though the mathematical results discussed here hold true for 
arbitrary (rational) $g>0$;\cite{BLL2} we use the subscript 0 to distinguish these bare fields from dressed boson fields $\hu(x)$ and 
$\hv(x)$ obtained from them by a Bogoliubov transformation, as described below.

The linear dynamics of these fields is given by the Hamiltonian (in this section and only here, we write $\int$ short for $\int_{-L/2}^{L/2}$, to simplify notation) 
\begin{equation} 
\label{cH2ell} 
\begin{split} 
\cH_2 = &\frac{g}{4\pi}\int\dd{x}\xxa \Big( \huz(x)^2+\hvz(x)^2  \\
& + \int \dd{x}' \Big[  U_2(x-x')
[\huz(x)\huz(x') \\ &+\hvz(x)\hvz(x')]  - U_1(x-x')\huz(x)\hvz(x')\Big]\Big) \xxe 
\end{split} 
\end{equation} 
with  colons indicating normal ordering and 
\begin{equation} 
\label{VW}
U_j(x) = \sum_{n=1}^\infty \frac{4q^{jn}}{1-q^{2n}}\cos(2\pi nx/L)\quad (j=1,2)
\end{equation} 
interaction potentials determined by the parameter
\begin{equation} 
q = \ee^{-2\pi\delta/L}\quad (\delta>0). 
\end{equation} 
The operator $\cH_2$ is a special case of a Luttinger Hamiltonian which, as is well-known, can be diagonalized by a Bogoliubov transformation.\cite{ML65}
This case is special in that the Bogoliubov transformed Hamiltonian has the same form as for $q=0$, except that the bare field operators are replaced by Bogoliubov transformed ones:\cite{L04} 
\begin{equation} 
\label{cH2diag} 
\cH_2 = \frac{g}{4\pi}\int\dd{x}\xxa \left(\hu^2+\hv^2 \right)\xxe. 
\end{equation} 
This is a consequence of the special form of the interactions in \eqref{VW}, and it corresponds to the fact that the Bogoliubov transformed fields $\hu$ and $\hv$ provide two commuting representations of the Virasoro algebra by the Sugawara construction, as in the special case $q=0$ where this is obvious; this is a manifestation of the fact that we are dealing with a nonchiral CFT (see e.g.\ Ref.~[\onlinecite{FMS97,G99}] for background on CFT). 
However, for nonzero $q$, the bare vacuum $|0\rangle$ is not a highest weight state for the dressed fields $\hu$ and $\hv$, and this has important consequences. 

The CFT described above accommodates the following two kinds of vertex operators,  
\begin{equation}
\label{phipmell}
\begin{split} 
\phi(x) &= \, \xxa \ee^{-\ii g\int^x \hu(x')\mathrm{d}x'}\xxe, \\
\tilde\phi(x) &= \, \xxa\ee^{\ii g\int^x\hv(x')\mathrm{d}x'}\xxe.
\end{split} 
\end{equation} 
Moreover, using the boson operators above, one can construct a self-adjoint operator, $\cH_3$, providing a second quantization of the elliptic CMS model in the following sense: 
this operator satisfies the relations
\begin{multline}
\label{L04} 
[\cH_3, \phi(x_1)\cdots  \phi(x_N)]|0\rangle \\ = H_N(\vx) \phi(x_1)\cdots  \phi(x_N)|0\rangle, 
\end{multline} 
for arbitrary particle number $N$. \cite{L00,L04}

We recently observed that it is possible to generalize $\cH_3$ so that one has   
relations similar to the ones in \eqref{L04} also for the vertex operators $\tilde\phi$.\cite{BLL3} 
This generalized operator can be written as 
\begin{multline} 
\label{cH3ell}
\cH_3 = \int:\bigg[\frac{g^2}{12\pi}\big( \hu^3+\hv^3\big)  + \frac{g(g-1)}{8\pi}\\
\times \big(  \hu T\hu_x + \hv T\hv_x 
+ \hu\tilde{T}\hv_x +\hv\tilde{T}\hu_x \bigr) \bigg]:\dd{x} 
\end{multline} 
with the integral operators $T$, $\tilde{T}$ in \eqref{TTe}--\eqref{MLe}.  
Thus, the operator $\cH_3$ defines the following quantum version of the periodic nonchiral ILW-equation, 
\begin{equation} 
\label{qILW} 
\begin{split} 
\hat{u}_t + 2\xxa\hat{u}\hat{u}_x\xxe + \frac12(g-1)[T\hat{u}_{xx}+\tilde{T}\hat{v}_{xx}]=0,\\
\hat{v}_t - 2\xxa\hat{v}\hat{v}_x\xxe - \frac12(g-1)[T\hat{v}_{xx}+\tilde{T}\hat{u}_{xx}]=0. 
\end{split}
\end{equation} 
To see this, we compute the Heisenberg equations of motion $A_t = \ii [\cH_3,A]$ for $A=\hu,\hv$ and rescale, $\hu\to \hat{u}\equiv g\hu/2$ and $\hv\to \hat{v}\equiv g\hv/2$, to obtain \eqref{qILW}. 
Moreover, by taking the classical limit where the boson operators $(\hat{u},\hat{v})$ become functions $(u,v)$ and $(g-1)$ is replaced by $g$, and specializing to $g=2$, \eqref{qILW}  reduces to \eqref{2ILW}.

It is interesting to note that the operator in \eqref{cH3ell}  satisfies the following generalization of \eqref{L04}, allowing for both kinds of vertex operators,  $\phi$ and $\tilde\phi$, at the same time: 
\newcommand{\tx}{\tilde{x}}
\begin{multline}
[\cH_3, \phi(x_1)\cdots  \phi(x_{N_1}) \tilde\phi(\tx_{1})\cdots  \tilde\phi(\tx_{N_2})]|0\rangle \\ = H_{N_1,N_2}(\vx,\tilde{\vx}) \phi(x_1)\cdots \tilde\phi(\tx_{N_2}) |0\rangle
\end{multline} 
where 
\begin{multline} 
\label{HNM}
H_{N_1,N_2}(\vx,\tilde{\vx}) = H_{N_1}(\vx) +H_{N_2}(\tilde{\vx}) \\
 + \sum_{j=1}^{N_1}\sum_{k=1}^{N_2}g(g-1)\wp_1(x_j-\tx_{k}+\ii\delta), 
\end{multline} 
 for arbitrary particle numbers $N_1,N_2$. 
This is a generalization of the elliptic CMS Hamiltonian \eqref{eCS} describing two types of particles, where particles of the same type interact with the singular two-body potential $\wp_1(x)$, and particles of different types interact with the nonsingular attractive potential $\wp_1(x+\ii\delta)$. 
It can be obtained from a standard elliptic CMS Hamiltonian \eqref{eCS} by dividing the particles into two groups and shifting the positions in one group by $\ii\delta$, similarly as in the classical case discussed in Section~\ref{sec:heuristic}; see \eqref{Newtonzw} {\em ff}.  
This argument proves that the Hamiltonian $H_{N_1,N_2}$ defines a quantum integrable model. However, the physically relevant eigenfunctions of $H_{N_1,N_2}$ can {\em not} be obtained from the ones of the corresponding  standard elliptic CMS Hamiltonian by this shift trick. This suggests that the generalized model can describe new physics which would be interesting to explore, but this is beyond the scope of the present paper. 

We finally mention that, to generate the full Hilbert space of the CFT, one needs to consider two further kinds of vertex operators representing hole excitations, and there is a generalization of the result in \eqref{HNM} allowing for arbitrary numbers, $N_1,M_1,N_2,M_2$, of all four types of vertex operators and with an interesting corresponding Hamiltonian $H_{N_1,M_1,N_2,M_2}$,\cite{BLL3} in generalization of a known result in the trigonometric case.\cite{AL17} 
Thus, $\cH_3$ is actually the second quantization of these operators  $H_{N_1,M_1,N_2,M_2}$ generalizing the elliptic CMS Hamiltonian. 

\subsection{Nonchiral CFT and FQHE} 
\label{sec:CFTB}
We motivate and explain our proposal that the nonchiral ILW equation can describe the interactions of nonlinear waves propagating on the two boundaries of a narrow FQHE system, in generalization of previous proposals for FQHE systems where the boundaries are well-separated and interboundary interactions can be ignored.\cite{W12} To prepare for this, we review known facts about the FQHE, bosonization and quantum hydrodynamics.

\subsubsection{Projection to lowest Landau level}
\label{sec:LLL}
We recall the quantum mechanical description of a charged particle confined to the $xy$-plane in the presence of a constant magnetic orthogonal to the plane (Landau problem): Assuming periodic boundary conditions in the $x$-direction: $-L/2\leq x\leq L/2$ with $L=L_x>0$, and $y\in\R$, the exact eigenfunctions in the lowest Landau level (LLL) have the form 
\begin{equation} 
\psi_k(x,y)= \ee^{\ii kx} \ee^{-(y-k)^2/2}, 
\end{equation} 
using the Landau gauge and units where the magnetic length is set to 1, with $k$ (short for $k_x$) an arbitrary integer multiple of $2\pi/L$. In such a state, the particle has the behavior of a plane-wave in the $x$-direction but is well-localized in the $y$-direction, and the quantum number $k$ therefore has a two-fold physical interpretation: it can be interpreted as momentum in $x$-direction and, at the same time, it corresponds to the location of the wave packet in $y$-direction. As is well-known, the wave functions in the LLL are all degenerate: the energy is $k$-independent. 

\begin{figure}[!htbp]
\centering
     \vspace{-.1cm}
\begin{overpic}[width=.42\textwidth]{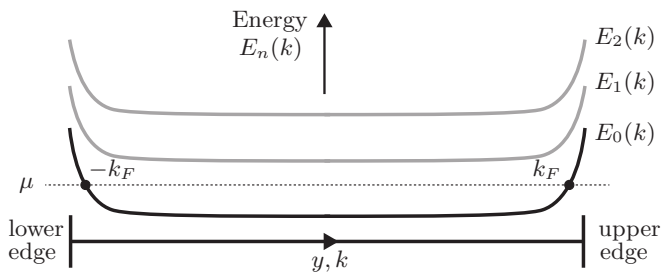}
      \put(47,0.5){\small $y,k$} 
      \put(33,43){\small Energy} 
      \put(34.2,38){\small $E_n(k)$} 
     \put(-6.5,6){\small lower} 
      \put(-6.5,2){\small edge} 
      \put(97.7,6){\small upper} 
      \put(98,2){\small edge} 
      \put(97,40){\small $E_2(k)$} 
      \put(97,32){\small $E_1(k)$} 
      \put(97,23){\small $E_0(k)$} 
      \put(-4.6,14.5){\small $\mu$} 
      \put(7.7,16.6){\small $-k_F$} 
      \put(86,16.6){\small $k_F$} 
     \end{overpic}
     \vspace{-.1cm}
\caption{Schematic picture of the lowest Landau level $E_0(k)$ in the presence of a potential confining the charged particles to a region $-L_y/2<y<L_y/2$, as illustrated in Fig.~\ref{Fig1}.
The grey lines indicate higher Landau levels that we ignore.}
\label{Fig1A} 
\end{figure}

We now consider the situation where, in addition to the magnetic field, we also have a potential, $\Vc(y)$, confining the charged particle to a region $-L_y/2\leq y\leq L_y/2$ for some $L_y>0$; this potential is zero at positions $y$ farther away than some distance $\ell_b>0$ from the boundary: $\Vc(y)=0$ for $|y\mp L_y/2|>\ell_b$, and it grows smoothly to very large values in the boundary regions $|y\mp L_y/2|<\ell_b$. In this situation, the degeneracy of the eigenfunctions in the LLL is lifted, and the energy $E_0(k)$ of the particle as a function of $k$ is qualitatively similar to the function $\Vc(y)$ with $y$ identified with $k$; see Fig.~\ref{Fig1A}. Thus, to describe noninteracting such particles projected to the LLL, one can use the quantum many-body Hamiltonian
\begin{equation} 
\label{HLLL} 
\cH_{\mathrm{LLL}} = \sum_{k} (E_0(k)-\mu) \hat\psi^\dag(k)\hat\psi(k)
\end{equation} 
with fermion field operators $\hat\psi^{(\dag)}(k)$ obeying canonical anticommutator relations,  $\{\hat\psi(k),\hat\psi^\dag(k')\}=\delta_{k,k'}$ etc.\ ($\mu$ is the chemical potential). 
By symmetry, we can assume $E_0(-k)=E_0(k)$. 

\vspace{-5pt}

Thus, even though we consider a two dimensional system, it is modelled by a one-dimensional Hamiltonian that can be treated by the bosonization method pioneered by Haldane.\cite{H81}
This bosonized description is useful since it allows to find interactions that can be added to the Hamiltonian without spoiling integrability; as discussed in the introduction, such interactions are particularly interesting in the context of FQHE physics. 

\subsubsection{Bosonization}
We recall some pertinent facts about bosonization.\cite{H81,LM15} 
Consider the free fermion model defined by the Hamiltonian \eqref{HLLL}. 
Its groundstate  is the Dirac sea where all states $-k_F<k<k_F$ are filled and all others are empty, with the Fermi momentum $k_F>0$ determined by $E_0(k_F)=0$. 
It is convenient to decompose the (inverse) Fourier transform of the fermion field, $\psi(x) = \sum_k (2\pi/L)\hat\psi(k)\ee^{\ii kx}$, as follows, 
\begin{equation} 
\label{psi} 
\psi(x) =  \psi_+(x)\ee^{\ii k_Fx} +  \psi_-(x)\ee^{-\ii k_Fx}
\end{equation} 
with fermion field operators $\psi_\pm(x)$ representing the low-energy excitations in the vicinity of the Fermi surface points $\pm k_F$. 
As explained in Section~\ref{sec:LLL}, these Fermi surface points can be identified with the two boundaries, $y=\pm L_y/2$,  of a FQHE system, as illustrated in Fig.~\ref{Fig1}. 

The fermion fields on the RHS in \eqref{psi} can be represented by vertex operators, 
\begin{equation}
\psi_\pm(x) = \, \xxa \ee^{\mp \ii \int^x \rho_\pm (x')\mathrm{d}x'}\xxe
\end{equation} 
where $\rho_\pm(x)$ are operators satisfying the commutator relations of chiral bosons,  
\begin{equation} 
[\rho_\pm(x),\rho_\pm(x')] = \mp 2\pi\ii\partial_x\delta(x-x')
\end{equation} 
and $[\rho_+(x),\rho_-(x')]=0$. 
These boson operators can be identified with the corresponding fermion densities, 
\begin{equation} 
\rho_\pm(x) = 2\pi \, \xxa \psi^\dag_\pm(x) \psi^\nd_\pm(x)\xxe. 
\end{equation} 
We note in passing that the boson fields $\rho_+(x)$ and $\rho_-(x)$ are equal, up to a factor $\sqrt{g}$ and zero mode details,\cite{AL17} 
to the bare boson fields $\huz(x)$ and $\hvz(x)$, respectively; see Section~\ref{sec:CFTA}. 

By Taylor-expanding the dispersion relation in the vicinity of the Fermi surface points: 
\begin{equation} 
E_0(\mp k_F + k)= \pm v_F k + \frac{k^2}{2m^*}+\ldots 
\end{equation} 
with the Fermi velocity $v_F=E_0'(k_F)$ and the effective mass $m^* = 1/E_0''(k_F)$, one can expand 
\begin{equation} 
\label{HLLL1} 
\cH_{\mathrm{LLL}} = v_F(\cH^{(0)}_{2,+}+ \cH^{(0)}_{2,-}) + \frac1{m^*}(\cH^{(0)}_{3,+}+ \cH^{(0)}_{3,-})+\ldots 
\end{equation} 
with 
\begin{equation} 
\label{cH023} 
\begin{split} 
\cH^{(0)}_{2,\pm} &= \frac1{4\pi}\int  \xxa \rho_\pm(x)^2\xxe \dd{x},\\
\cH^{(0)}_{3,\pm} &= \frac1{12\pi}\int \xxa \rho_\pm(x)^3 \xxe \dd{x}  
\end{split} 
\end{equation} 
etc.  This provides a basis for the quantum hydrodynamic description of such systems proposed by Abanov and Wiegmann.\cite{AW05} 

\subsubsection{Chiral Luttinger liquids and FQHE} 
We recall Wen's chiral Luttinger liquid description of FQHE systems.\cite{W90} 

The leading term in \eqref{HLLL1},  
\begin{equation} 
\cH^{(0)}_2=v_F(\cH^{(0)}_{2,+}+\cH^{(0)}_{2,-}), 
\end{equation} 
provides a good starting point to describe FQHE systems, but the low-energy excitations are not fermions but rather collective excitations that can be described by vertex operators 
\begin{equation} 
\label{psipmg}
\phi_\pm(x) = \, \xxa \ee^{\mp \ii \sqrt{g} \int^x \rho_\pm (x')\mathrm{d}x'}\xxe
\end{equation} 
with $g=3,5,\ldots$ at filling level $\nu=1/g$; the fermion case $g=1$ corresponds to the integer Hall effect and, for $g>1$, the vertex operators \eqref{psipmg} describe composite fermions.  
If the two boundaries are far apart, it is natural to assume that the low-energy excitations at distinct boundaries do not interact, and one can restrict the discussion to {\em one} boundary or, equivalently, 
to {\em one} chiral sector, $+$ or $-$. This is Wen's chiral Luttinger liquid model.\cite{W90} 

\subsubsection{Boundary waves in FQHE systems}
\label{sec:wave}
The dynamics of the boson fields provided by the Hamiltonian $\cH^{(0)}_2$ via the Heisenberg equations of motion is 
\begin{equation} 
\label{wavepm} 
\partial_t\rho_\pm \pm v_F \partial_x \rho_\pm =0. 
\end{equation} 
These linear equations describe waves propagating at the two boundaries of a FQHE system:\cite{W12} at each boundary, the wave packets move in one direction, right $(+)$ or left $(-)$, 
with constant speed $v_F$ and without changing shape. 

The Hamiltonian $\cH^{(0)}_2$ is highly degenerate, and it is natural to ask if one can lift this degeneracy by adding interactions that fulfill the following requirements: (i) they do not spoil integrability, (ii) they provide nonlinear corrections to the linear wave equations \eqref{wavepm}, (iii) they are compatible with the vertex operators \eqref{psipmg} describing composite fermions.\cite{W12} 
An interesting Hamiltonian obtained by adding such terms to $\cH^{(0)}_{3,\pm}$ \eqref{cH023} is\cite{AW05}  
\begin{equation} 
\label{cH3} 
\cH_{3,\pm} = \int \xxa\Bigl(  \frac{\sqrt{g}}{12\pi}\rho_\pm(x)^3 + \frac{g-1}{8\pi}\rho_\pm H(\rho_\pm)_x \Bigr)  \xxe \dd{x}  
\end{equation}  
with the Hilbert transform $H$ (obtained from $T$ in \eqref{TTe} by taking the limit $\delta\to\infty$): the dynamics for the boson fields provided by this Hamiltonian is a quantum version of the BO equation which is integrable,\cite{AW05} 
and this Hamiltonian is compatible with the composite fermion operators in that it also provides a second quantization of trigonometric CMS model;\cite{AI94,I95,AMOS95,ES98,CL99,BH08,AL17}  
using the latter and the  known eigenfunctions of the trigonometric CMS system, one can construct the exact eigenstates of $\cH_{3,\pm}$.\cite{AL17}

\subsubsection{Proposal}
\label{sec:proposal} 
We now are ready to motivate and explain our proposal that the nonchiral ILW equation can describe waves propagating on parallel boundaries of FQHE systems.

We recall that, in generic applications of bosonization, the most important interactions to be added to the Hamiltonian $\cH^{(0)}_2$ are quadratic in the boson operator, and thus, 
generically, one obtains a Luttinger Hamiltonian as in \eqref{cH2ell}, for some potentials, $U_2(x)$ and $U_1(x)$;\cite{H81} these potentials describe interactions between the same ($U_2$) and opposite ($U_1$) chiral degrees of freedom. 
Moreover, one often assumes that these interactions are local since this guarantees that the resulting model is conformally invariant. 
In the context of the FQHE, such Luttinger interactions are usually ignored by the following arguments: (i) the two chiral degrees of freedom describe excitations at two separated boundaries of the system, and $U_2(x)$ therefore describes interedge interactions which are negligible if the boundaries are sufficiently far apart; (ii) a local interaction $U_1(x)$ within the same boundary only renormalizes the Fermi velocity and thus can be taken into account by redefining $v_F$. 
However, since the one-particle eigenfunctions of the Landau Hamiltonian are spatially extended, and Coulomb interactions in a FQHE system are long-range, there is no reason to exclude nonlocal interactions which preserve conformal symmetry. 
Moreover, it is known that transport coefficients in Luttinger liquid are universal even if the interactions mix the chiral degrees of freedom and are nonlocal,\cite{LLMM17} i.e., the accurate quantization of the Hall conductance observed in real FQHE systems is compatible with generic Luttinger model interactions; see Ref.~[\onlinecite{FM20}] for a recent construction of the pertinent general Luttinger model for general vertex operators as in \eqref{psipmg}.

As discussed in Section~\ref{sec:CFTA}, the Luttinger Hamiltonian \eqref{cH2ell} with the fine-tuned interactions in \eqref{VW} is conformally invariant, for arbitrary fixed $\delta>0$, and there are natural corresponding generalizations of the composite fermion operators and the operator in \eqref{cH3} satisfying the requirements stated in Section~\ref{sec:wave}: they are given in \eqref{phipmell} and \eqref{cH3ell}, respectively. Our proposal to model a FQHE system at filling $1/g$, $g=3,5,\ldots$, and with the geometry illustrated in Fig.~\ref{Fig1} is therefore as follows: {\em The boson field operators $\huz$ and $\hvz$, satisfying the commutator relations in \eqref{CCR0}, describe low-energy excitations located at the upper and lower edge, respectively,  of the FQHE system boundary; the low-energy description of the system is by the Hamiltonian $\cH_2$ in \eqref{cH2ell}--\eqref{VW}, with the parameters $v_F$ and $\delta$ determined by system details like the edge distance, $L_y$, and the confining potential, $\Vc(y)$; the linear- and nonlinear dynamics of the boundary waves is described by the operators in \eqref{cH2ell} and \eqref{cH3ell}, respectively; the vertex operators in \eqref{phipmell} describe quasiparticle excitations in the system.}

\subsubsection{Inter-edge effects in FQHE systems} 
We argue that the model proposed in Section~\ref{sec:proposal} can describe interedge effects in narrow FQHE system. 

Our model predicts that the quasiparticles of the system are the Bogoliubov transformed boson fields, $\hu$ and $\hv$, diagonalizing $\cH_2$ in \eqref{cH2ell}; see \eqref{cH2diag}. Thus, the linear dynamics is given by the same equations as for $\delta=\infty$, i.e.,  $\hu_t-v_F\hu_x=0$ and $\hv_t+v_F\hv_x=0$. However, since $\hu$ (say\cite{remsimilar}) is a superpositions of the fields  $\huz$ and $\hvz$ localized at two distinct boundaries, a right-moving wave excited at the upper edge will always develop into a pair of well-defined corresponding excitations at both edges moving in parallel. Thus, our proposal can be tested already in experiments on real FQHE systems that can only resolve {\em linear} boundary waves: our model predicts corresponding excitations, $u_0(x-v_Ft)$ and $v_0(x-v_Ft)$ proportional to the expectation values of $\huz(x,t)$ and $\hvz(x,t)$, respectively, where $u_0(x)$ and $v_0(x)$ are determined by a single function, $u(x)$, and the inverse of the Bogoliubov transformation described in Section~\ref{sec:CFTA} ($u(x)$ is proportional to the expectation value of $\rho(x,t=0)$). Furthermore, our model predicts nonlinear waves described by the quantum nonchiral ILW equation \eqref{qILW}, in generalization of the Wiegmann proposal  quoted in the beginning of this section.\cite{W12}  It would be interesting to elaborate these predictions in detail, and to propose specific experiments on real FQHE systems to test them. Clearly, this is a research project in its own. Our results in Sections~\ref{sec:hyp} and \eqref{sec:ell} are a first step, giving an indication of the new physics that the nonchiral ILW equation can describe.

To elaborate predictions of our model, it would be interesting to construct the exact eigenstates of the Hamiltonian $\cH_3$ in \eqref{cH3ell}, in generalization of known results for $\delta=\infty$.\cite{AL17} 
This is challenging. One reason is that, while the exact eigenstates of the trigonometric CMS model have been known for a long time, 
the ones of the relevant elliptic CMS-type systems are the subject of ongoing research.\cite{AL20}

\section{Results: Hyperbolic Case} 
\label{sec:hyp} 
\subsection{Multisoliton solutions} 
The following fundamental result shows that (\ref{2ILW}) admits multisoliton solutions whose dynamics is  described by the hyperbolic CMS model, thus generalizing a famous result for the rational case:\cite{CLP79} {\em For arbitrary integers $N\geq 1$ and complex parameters  $a_{j}$ with imaginary parts in the range $\delta/2<\Im a_{j}<3\delta/2$ for $j=1,\ldots,N$, the following is an exact solution of the nonchiral ILW equation 
 \eqref{2ILW}:
\begin{equation} 
\label{solitonN} 
 \left(\begin{array}{c} u(x,t) \\ v(x,t)  \end{array}\right)  =  \ii\sum_{j=1}^N\left(\begin{array}{c} \alpha(x-z_j(t)-\ii\delta/2)\\  -\alpha(x-z_j(t)+\ii\delta/2)   \end{array}\right) + \mathrm{c.c.}
\end{equation}  
where $\alpha(x)= (\pi/2\delta)\coth(\pi x/2\delta)$ and the poles $z_j(t)$ are determined by Newton's equations \eqref{Newton} with $V(r)$ given by \eqref{V} and with initial conditions $z_j(0)=a_{j}$ and} 
\begin{equation} 
\label{dotz} 
\dot z_j(0) = 2\ii\sum_{j'\neq j}^N\alpha(a_{j}-a_{j'}) -2\ii \sum_{k=1}^N\alpha(a_{j}-\overline{a}_{k}+\ii\delta) 
\end{equation} 
(the bar denotes complex conjugation, c.c.). 
Thus, to obtain an exact solution of \eqref{2ILW}, one chooses complex parameters $a_j$ satisfying $\delta/2<\Im a_j<3\delta/2$; next, the time-evolution of $z_j(t)$ is obtained by solving the hyperbolic CMS model with initial conditions determined by the $a_j$; finally, the solution of (\ref{2ILW}) is obtained from (\ref{solitonN}).
Using the exact analytic solution of the hyperbolic CMS model obtained by the projection method,\cite{OP81} the numerical effort to compute such an multisoliton solution at an arbitrary time, $t$,  is reduced to diagonalizing an explicitly known $N\times N$ matrix. As elaborated in Appendix~\ref{app:B}, we tested this result by comparing with a numeric solution of \eqref{2ILW}. 

\vspace*{-\baselineskip}
\onecolumngrid
\begin{widetext} 
\begin{figure}[!htbp]
\centering
\begin{overpic}[scale=.21]{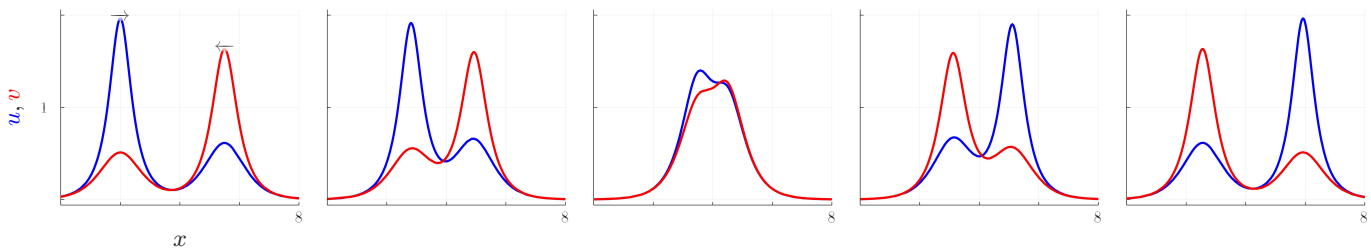}
      \put(-1.0,7.5){\small \begin{rotate}{90}${\color{blue} u} , {\color{red} v}$\end{rotate}}
      \put(10.5,-1.5){\small $x$}
     \end{overpic}
\caption{
Time evolution of a two-soliton solution of the nonchiral ILW equation \eqref{2ILW} with a $u$-channel dominated soliton (big blue and small red humps) colliding with a $v$-channel dominated soliton (big red and small blue humps),
as explained in the main text. 
The plots show $u(x,t)$ (blue line) and $v(x,t)$ (red line) at successive times $t=(n-1)t_0$, $n=1,\ldots,5$; the parameters are $\delta=\pi$, $a_1=-4+1.2\ii\delta$, $a_2=3+0.85\ii \delta$, and $t_0=2.25$. 
}     
\label{Fig2} 
\end{figure}
\vspace{-5pt}
\end{widetext}
\twocolumngrid

\begin{figure}[!htbp]
\centering
\vspace{-2pt}
\begin{overpic}[scale=.21]{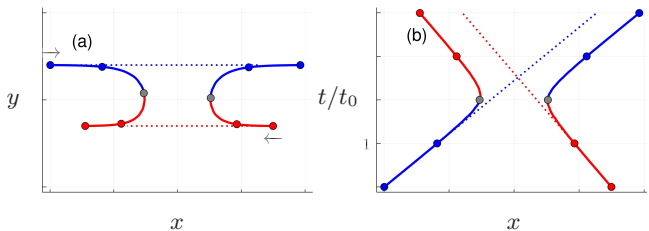}
      \put(-0.5,16.7){\small $y$} 
      \put(46.0,16.7){\small $t/t_0$} 
      \put(23.9,-2.0){\small $x$} 
      \put(74.25,-2.0){\small $x$} 
     \end{overpic}
\caption{
(a) Time evolution of the poles  $z_j(t)$, $j=1,2$, in the complex plane corresponding to the two-soliton solution in Fig.~\ref{Fig2}. 
The times $t=(n-1)t_0$, $n=1,\ldots,5$, defined in the caption of Fig.~\ref{Fig2} are indicated by circles; the arrows mark circles corresponding to $n=1$.
The dotted lines indicate the evolution of poles without interactions.
(b) Time evolution of the center-of-mass locations of the solitons given by $\Re z_j(t)$.}
\label{Fig3}
\end{figure}
\vspace{-5pt}

\subsection{Examples.}The one-soliton solution of (\ref{2ILW}) is given by 
\begin{equation} 
\label{soliton1} 
 \left(\begin{array}{c} u(x,t) \\ v(x,t)  \end{array}\right)  =  \ii \left(\begin{array}{c} \alpha(x-z(t)-\ii\delta/2)\\  -\alpha(x-z(t)+\ii\delta/2)   \end{array}\right) + \mathrm{c.c.} , 
\end{equation} 
where the poles evolve linearly in time,  with initial conditions determined by a complex parameter $a$ such that $\delta/2<\Im a\leq 3\delta/2$, 
\begin{equation} 
z(t) = a+\dot z(0)t, \quad \dot{z}(0)=2\ii\alpha(a-\bar{a}+\ii\delta). 
\end{equation} 
It is important to note that $\dot z(0)$ is real, and therefore, $\Im z(t)=\Im a$ independent of $t$. Thus, the functions $u(x,t)$ and $v(x,t)$ both describe humps whose shapes do not change with time. 
These humps are centered at the same point and move with constant velocity, $\Re z(t)=\Re a + \dot{z}(0) t$, and their heights, $\max u>0$ and $\max v>0$, are determined by $\Im a$. 
For $\Im a$ close to $3\delta/2$, $\max u\gg \max v$, and the solitons move to the right, $\dot z(0)>0$. 
As $\Im a$ decreases, $\max u$ and $\dot z(0)$ decrease while $\max v$ increases until, at $\Im a=\delta$, $\max u=\max v$ and $\dot z(0)=0$. 
Thus, if  $\Im a$ lies in the range $\delta <\Im a < 3\delta/2$, then the one-soliton is mainly in the $u$-channel and moves to the right; it is therefore similar to the one-soliton solution of the standard ILW equation $u_t+2uu_x+Tu_{xx}=0$. Similarly, when $\delta/2 <\Im a < \delta$, the one-soliton is mainly in the $v$-channel and moves to the left, similar to a one-soliton solution of the $P$-transformed  ILW equation $v_t - 2 vv_x-Tv_{xx}=0$. 

For parameters $a_j$ such that $\Re (a_j-a_k)\gg \delta$ for all $j\neq k$, the multisoliton solution of (\ref{2ILW}) is well-approximated by a sum of $N$ one-solitons  of the form \eqref{soliton1} where $\dot z_j(t) \approx 2\ii\alpha(a_j-\bar{a}_j+\ii\delta)$ is time-independent for times such that $\Re (z_j(t)-z_k(t))\gg \delta$;  see Fig.~\ref{Fig2} for a two-soliton solution, with the corresponding motion of poles in Fig.~\ref{Fig3}(a).
However, when two solitons meet, they interact in a nontrivial way, and after the interaction they re-emerge with the same shape but with phase-shifts; see Fig.~\ref{Fig3}(b). 
Such nontrivial interactions between solitons can also be modeled by the system of decoupled ILW equations obtained from \eqref{2ILW} by dropping the $\tilde{T}$-terms. 
A qualitatively new effect stemming from the $\tilde{T}$-terms is that $u$-channel dominated solitons ($u$-solitons) interact nontrivially with $v$-solitons, as clearly seen in our example in Figs.~\ref{Fig2} and \ref{Fig3}.
It is interesting to note that the poles corresponding to the $u$- and $v$-solitons interchange their imaginary parts and directions during the collision and thus, in this sense, exchange their identities: while the first pole corresponds to the $u$-soliton and the second to the $v$-soliton before the collision, it is the other way round after the collision; see Figs.~\ref{Fig3}(a) and (b).
We note that such an identity change of poles during soliton collisions is known for the BO equation,\cite{M80} but only for solitons moving in one direction. 

\subsection{Derivation of multisoliton solutions.}
We explain the key difference between the derivation of solitons for (\ref{2ILW}) and the corresponding derivation in the rational case;\cite{CLP79} further details can be found in Appendix~\ref{app:hyp}. 

The Hilbert transform, $H$, satisfies $H^2=-I$, and this property is crucial for the existence of eigenfunctions of $H$ needed in the derivation of the CMS-related soliton solutions of the BO equation $u_t+2uu_x+Hu_{xx}=0$.\cite{CLP79}
However, while the trigonometric generalization of $H$ also has this property, the hyperbolic generalization of $H$ is the operator $T$ in \eqref{TT}, and $T^2\neq -I$. 
This is the reason why the soliton solution of the BO equation straightforwardly generalizes to the trigonometric case,\cite{CLP79} but the naive generalization to the hyperbolic case fails. 
However, the nonchiral ILW equation \eqref{2ILW} can be written in vector form as
\begin{equation} 
\label{2ILW2} 
\begin{split} 
&\vu_t +  (\vu.\vu)_x + \cT \vu_{xx}=0, \\
\vu\equiv \left(\begin{array}{c} u \\ v  \end{array}\right)\!,\;\,& \vu.\vu \equiv \left(\begin{array}{c} u^2 \\- v^2  \end{array}\right)\!,\;\,
\cT \equiv \left(\begin{array}{cc} T & \tilde{T} \\ -\tilde{T} & -T \end{array}\right)  
\end{split}
\end{equation} 
where the matrix operator, $\cT$, satisfies $\cT^2=-I$. Moreover, $(\alpha(x+z\pm \ii\delta/2),-\alpha(x+z\mp\ii\delta/2))^t$ are eigenfunctions of $\cT$ with eigenvalues $\pm \ii$. 
The latter are the eigenfunctions needed to be able to use the method developed for the rational case:\cite{CLP79}  using well-known identities for the function $\alpha(x)$,\cite{C75} as well as a B\"acklund transformation for the hyperbolic CMS model,\cite{W82} it is straightforward to adapt a known derivation of multisoliton solutions of the BO equation\cite{CLP79} to the hyperbolic case.

\subsection{Integrability.}
We found a Lax pair, a Hirota bilinear form, a B\"acklund transformation, and an infinite number of conservation laws for \eqref{2ILW}. 
Thus, the nonchiral ILW equation is a soliton equation that is integrable in the same strong sense as the standard ILW equation.\cite{KSA81} 
Below we present some of these results that can be checked by straightforward computations.

The Lax pair we found is as follows:
{\em Let $\psi(z;t,k)$ be an analytic  function  on the union of the strips $0<\Im z<\delta$ and $\delta<\Im z<2\delta$ and extended to $\C$ by $2\ii\delta$-periodicity, $\psi_0^{\pm}(x;t,k)$ and $\psi_\delta^{\pm}(x;t,k)$ the boundary values of this function on $\R$ and $\R+\ii\delta$, respectively, and $\mu_1$, $\mu_2$, $\nu_1$, and $\nu_2$ arbitrary functions of the spectral parameter $k$. 
Then the compatibility of the following linear equations yields \eqref{2ILW}:  
\begin{equation*} 
\begin{split} 
&(\ii\partial_x -u-\mu_1)\psi_{0}^- = \nu_1 \psi_0^+,\quad 
(\ii\partial_x+v-\mu_1) \psi_{\delta,x}^+  = \nu_2 \psi_\delta^-, 
	\\
&\left(\ii\partial_t- 2\mu_1\ii \partial_x -\partial_x^2 + Tu_x + \tilde{T}v_x \pm \ii u_x + \mu_2 \right) \psi_0^\pm=0, 
	\\
&\left( \ii\partial_t - 2\mu_1\ii\partial_x -\partial_x^2  +Tv_x  + \tilde{T}u_x  \pm \ii v_x + \mu_2\right)  \psi_{\delta}^\pm =0 .
\end{split} 
\end{equation*} 
}

Inspired by known results for the BO equation,\cite{ABW09} we obtained the following Hirota bilinear form of (\ref{2ILW}),
\begin{equation}
(\ii D_t-D_x^2)F^{-}\cdot G^{+} = (\ii D_t-D_x^2)F^{+}\cdot G^{-}=0
\end{equation}
with $u=\ii\partial_x\log (F^-/G^+)$ and $v=-\ii\partial_x\log (F^+/G^{-})$, where $F^\pm(x,t)\equiv F(x\pm \ii\delta/2,t)$ and similarly for $G$, using standard Hirota derivatives.\cite{HirotaD}

The first three of the conservation laws we found are
\begin{equation}\label{conservationlaws}
\begin{split} 
I_1=&\int_{\R}(u+v)\mathrm{d}x,\quad I_2=\frac12\int_{\R}(u^2-v^2)\mathrm{d}x, \\
I_3=&\int_{\R}\bigg[\frac{u^3}3+\frac{uTu_x}{2} + \frac{u\tilde{T}v_x}{2} + (u\leftrightarrow v) \bigg]\mathrm{d}x
\end{split} 
\end{equation}
with $(u\leftrightarrow v)$ short for the same three terms but with $u$ and $v$ interchanged.

B\"acklund transformations, other conservation laws, and detailed derivations are given elsewhere.\cite{BLL2}

\section{Results: Elliptic case}
\label{sec:ell}
To generalize (\ref{2ILW}) to the periodic setting, we use the Weierstrass functions $\wp(z)$ and $\zeta(z)$ with periods $(2\omega_1,2\omega_2)\equiv (L,2\ii\delta)$,\cite{WW40} $L>0$, and the related functions $\zeta_j(z)\equiv \zeta(z)-\eta_j z/\omega_j$, $\eta_j\equiv \zeta(\omega_j)$,  $j=1,2$.
The function $\zeta_1(z)$ is $L$-periodic, $\zeta_1(z+L)=\zeta_1(z)$, whereas the function $\zeta_2(z)$ is $2\ii\delta$-periodic, $\zeta_2(z+2\ii\delta)=\zeta_2(z)$; recall that $\zeta(z)$ is neither $L$- nor $2\ii\delta$-periodic.
We note that $\wp_1(x)$ in \eqref{wp1} equals $-\zeta_1'(x)=\wp(x)+\eta_1/\omega_1$.

The periodic nonchiral ILW equation is given by \eqref{2ILW} with the integral operators  $T$, $\tilde{T}$ in \eqref{TTe}--\eqref{MLe}.
With that, $\cT$ in \eqref{2ILW2} still satisfies $\cT^2=-I$,  and the derivation of the multisoliton equation outlined above generalizes straightforwardly to the elliptic case provided $\alpha(z)$ in \eqref{ansatz} is chosen as the $2\ii\delta$-periodic variant of $\zeta(z)$: {\em The functions $u(x,t)$ and $v(x,t)$ given in \eqref{solitonN}, with $\alpha(x)=\zeta_2(z)$,  satisfy the periodic nonchiral ILW equation provided that $z_j(t)$ satisfy Newton's equations \eqref{Newton} with the elliptic CMS model potential $V(r)=\wp(r)$, and with initial conditions $z_j(0)=a_j$ and $\dot z_j(0)$ in \eqref{dotz}, for arbitrary complex $a_j$ satisfying $\delta/2<\Im a<3\delta/2$ and $-L/2\leq \Re a_j<L/2$, $j=1,\ldots,N$.}
It is important to note that the multisoliton solution is $L$-periodic even though $\zeta_2(z)$ is not.
The interested reader can find further details in Appendix~\ref{app:ell}.

\section{Other applications} 
\label{sec:applications}
We present arguments suggesting that the nonchiral ILW equation introduced in this paper will find other applications in physics beyond the application to the FQHE described earlier (Section~\ref{sec:applA}). 
As a specific example, we discuss a possible application in the context of nonlinear water waves, and thereby provide a complementary physical interpretation of our mathematical results (Section~\ref{sec:applB}). 

\subsection{The wide applicability of soliton equations} 
\label{sec:applA}
Nonlinear evolution equations are typically more difficult to solve than linear ones, and theoretical physics tools are often not equally powerful when nonlinear effects are important. 
Soliton equations are an important exception: these nonlinear equations are integrable, and it is therefore possible to develop analytic \cite{AC91}  and numeric \cite{AH87} methods to solve them reliably.
Thus, phenomena described by soliton equations can be very well understood despite of the crucial importance of nonlinear effects. 
The class of such phenomena is remarkably large, with many examples from different areas in physics such as hydrodynamics, nonlinear optics, plasma physics, dislocation theory of crystals, etc. 
A well-known explanation of this wide applicability of soliton equations is by Calogero:\cite{C91}
{\em certain ``universal" nonlinear PDEs \cite{remPDE} can be obtained, by a
limiting procedure involving rescalings and an asymptotic expansion, from very
large classes of nonlinear evolution equations [\ldots]. Because this
limiting procedure is the correct one to evince nonlinear effects, the universal
model equations obtained in this manner [\ldots] are widely applicable.
Because this limiting procedure
generally preserves integrability, these universal model equations are likely to
be integrable [\ldots].} 

This suggests that the nonchiral ILW \eqref{2ILW} will find other applications in physics.

\subsection{Nonlinear water waves}
\label{sec:applB}
Consider the following class of soliton equations describing, e.g., nonlinear water waves in different situations: 
\begin{equation}
\label{ueq} 
u_t +2 u u_x + Du_{xx}  = 0 
\end{equation} 
where $D$ is one of the linear operators specified below and $u= u(x,t)$, where $x$ is a coordinate on one-dimensional space and $t$ time.
This class includes the famous {\em Korteweg-de Vries} (KdV) equation,\cite{DJ89} the BO equation,\cite{B67,DA67,O75} the ILW equation interpolating between the KdV and the BO equations,\cite{J77} and periodic variants of these three equations \cite{AFSS82} depending on a further parameter, $L>0$, corresponding to the spatial period: $u(x+L,t)=u(x,t)$.

While the nonlinear term, $2 uu_x$, is the same in all cases, the dispersive term, $Du_{xx}$, is different: it amounts to multiplication of $u$ by functions $\ii\Om(k)$ in Fourier space:  $Du_{xx}=\ii\Om(-\ii\partial_x)u$,  with the following dispersion relations in the different cases,\cite{remILW}  
 \begin{equation} 
 \label{Om0} 
 \Om(k) = \begin{cases} k^3\delta/3 & \text{(KdV)} \\ 
 k^2\sign(k) & \text{(BO)} \\
 k^2\coth(k\delta) & \text{(ILW)} \end{cases}
 \end{equation} 
where the wave number, $k$,  is restricted to integer multiples of $2\pi/L$ in the periodic cases (it is real otherwise), and $\delta>0$ is a constant. 
Note that, in position space, the operator $D$ is represented by a differential operator in the KdV case, $D f=\delta \partial_x f/3$, whereas in the BO- and ILW cases it is given by an integral operator denoted as $H$ (Hilbert transform) and $T$, respectively; see \eqref{hatTT}. 
 
It is important to note that, in general, one should add a term $c u_x$ to the LHS in \eqref{ueq}, with $c$ some velocity parameter, to make manifest that \eqref{ueq} is a generalization of the chiral wave equation $u_t+cu_x=0$; however, since this term is trivial in that it can be removed by a transformation $u\to u-c/2$, we ignore it in our discussion. 

The soliton equations in \eqref{ueq}--\eqref{Om0} provide effective descriptions of nonlinear water waves taking into account the most important nonlinear and dissipative terms.\cite{J77}  
It is important to note that, when deriving these equations from fundamental hydrodynamic laws, parity invariance is broken and, for this reason, the equations in \eqref{ueq}--\eqref{Om0} are {\em chiral}: they can only describe solitons moving to the right. 
Obviously, one can obtain a corresponding equation describing solitons moving to the left by a parity transformation: $v(x,t)\equiv u(-x,t)$ satisfies $v_t-2vv_x-Dv_{xx}=0$. 
Thus, the chiral equation in \eqref{ueq}--\eqref{Om0} actually corresponds to a system of two equations for $u$ and $v$ describing solitons moving in both directions. 

Clearly, this description is simplistic with regard to the following: solitary waves in nature moving in opposite directions interact when they meet, but such interactions are ignored by this uncoupled system for $u$ and $v$.
This suggests to try to find integrable generalizations of these equations of the form 
\begin{equation} 
\label{uveq} 
\begin{split} 
u_t +2 u u_x + Du_{xx} +X(v,u) &= 0,  \\
v_t -2 v v_x - Dv_{xx}  -X(u,v) & = 0 
\end{split} 
\end{equation}
with coupling terms,  $X(v,u)$ and $X(u,v)$, such that the system \eqref{uveq} is invariant under the parity transformation in \eqref{P}.
We believe that neither the KdV equation  nor the BO equation allow for such a coupling; however, the ILW equation does:
it is given by the dispersive term $I(v,u)=\tilde{D}v_{xx}=\ii\tilde{\Om}(-\ii\partial_x)v$ (independent of $u$) with
\begin{equation}
 \tilde{\Om}(k)=\frac{k^2}{\sinh(k\delta)};
\end{equation}
indeed, using \eqref{hatTT}, one sees that \eqref{uveq} in this case is equivalent to the nonchiral ILW equation \eqref{2ILW} {\em ff}.

One can check that \eqref{2ILW} does not have a well-defined limit $\delta\to 0$, and that $\tilde{T}u_{xx}\to 0$ in the limit $\delta\to \infty$: the KdV-limit of the nonchiral ILW equation does not exist, and its BO-limit is trivial.
Thus, to describe nonchiral physics, one has to work in the regime $0<\delta<\infty$.

This suggests that it would be interesting to revisit the derivation of the KdV-equation from more fundamental parity invariant equations, and to see if this can be generalized so as to obtain the nonchiral ILW equation.

\section{Final remarks}
\label{sec:final}
We presented the novel soliton equation \eqref{2ILW}. 
We call it the nonchiral ILW equation because it is parity invariant and can describe interacting solitons moving in both directions. 
We obtained exact multisoliton solutions determined by poles satisfying the equations of motion of the hyperbolic CMS model, and we gave a Lax pair, a Hirota form, and conservation laws.  
We also presented a periodic nonchiral ILW equation and its soliton solutions determined by the elliptic CMS model.

We proposed that the nonchiral ILW equation can model coupled nonlinear waves in FQHE systems, and we gave background information to make this proposal precise. 
However, as we argued, our results are of wider interest: 
Many soliton equations containing only first-order derivatives in time are chiral, i.e., they can only describe solitons moving in one direction, left or right, and thus are not parity invariant. 
Examples include the KdV equation, the BO equation and, more generally, the ILW equation. 
However, the fundamental equations in hydrodynamics from which these soliton equations are derived are parity invariant. 
This mismatch of symmetries is not fully satisfactory. 
Using the nonchiral ILW equation instead of the standard ILW equation reconciles symmetries, and 
 we therefore believe that, in various applications in physics,  the former can be a better approximation to fundamental equations than the latter.

We hope that our results open up a route to generalize recent results on a generalized hydrodynamic description of the Toda chain\cite{S19,D19} to the elliptic CMS model. 
This would be interesting  since, in the elliptic CMS model, one can change the qualitative character of the interaction from long-range in the trigonometric case, to short-range in the hyperbolic case, to nearest-neighbor in the Toda limit.

\begin{acknowledgments} 
We thank L. Bystricky,  M. Noumi, and  in particular J. Shiraishi for very helpful and inspiring discussions.
B.K.B.\ acknowledges support from the G\"oran Gustafsson Foundation and the European Research Council, Grant Agreement No. 682537.
E.L.\ acknowledges support by the Swedish Research Council, Grant No.\ 2016-05167, and by the Stiftelse Olle Engkvist Byggm\"astare, Contract 184-0573.
J.L. acknowledges support from the G\"oran Gustafsson Foundation, the Ruth and Nils-Erik Stenb\"ack Foundation, the Swedish Research Council, Grant No.\ 2015-05430, and the European Research Council, Grant Agreement No. 682537.
\end{acknowledgments} 

\appendix 
\section{Derivation of soliton solutions}
\label{app:A} 
We give details on the derivation of the $N$-soliton solutions presented in the main text, both in the hyperbolic and elliptic cases. 

\subsection{Hyperbolic case}
\label{app:hyp} 

We construct solutions of \eqref{2ILW2} with $T,\tilde{T}$ defined in \eqref{TT} by generalizing a known method for the BO equation.\cite{CLP79,SAX08}

\subsubsection{Integral operators in Fourier space} 
We compute the Fourier space representation of the matrix operator $\cT$ in \eqref{2ILW2}.

We start by transforming the operators $T,\tilde{T}$ in \eqref{TT} to Fourier space, using the following exact integral, 
\begin{equation} 
\label{J} 
\int_{\R}{} \frac{\pi}{2\delta}\coth\left(\frac{\pi }{2\delta}(x\mp \ii a) \right)\ee^{-\ii k x}\dd{x} =  -\pi\ii\frac{\ee^{\pm(ak-k\delta)}}{\sinh(k\delta)}
\end{equation} 
for real parameters $a, k$ such that $0<a<2\delta$ and $k\neq 0$ (a derivation of this result can be found at the end of this section). 
This implies 
\begin{equation} 
\label{FT}
\begin{split} 
\fpint{\R}{} \frac{1}{2\delta}\coth\left(\frac{\pi }{2\delta}x\right)\ee^{-\ii k x}\dd{x}&= -\ii\coth(k\delta),\\ 
\int_{\R}{} \frac{1}{2\delta}\tanh\left(\frac{\pi }{2\delta}x\right)\ee^{-\ii k x}\dd{x}&= -\ii\frac{1}{\sinh(k\delta)}
\end{split} 
\end{equation} 
for real $k\neq 0$. Indeed, the first of these identities is equivalent to the average of the two integrals in \eqref{J} in the limit $a\downarrow 0$, and the second is obtained from \eqref{J} in the special case $a=\delta$. 
Observe that the integrals in \eqref{TT} are convolutions. 
Using the following conventions for Fourier transformation, $\hat u(k)=\int_{\R} u(x)\ee^{-\ii kx}\dd{x}$, the operators defined in \eqref{TT} can therefore be expressed in Fourier space as follows,  
\begin{equation} 
\label{hatTT} 
\begin{split} 
&\widehat{(Tu)}(k) = \ii\coth(k\delta)\hat u(k),\\
&\widehat{(\tilde{T}u)}(k)  = \ii\frac1{\sinh(k\delta)}\hat u(k). 
\end{split} 
\end{equation} 
Thus, for the matrix operator $\cT$ defined in \eqref{2ILW2}, $\widehat{\cT\vu}(k)=\hat{\cT}(k)\hat{\vu}(k)$ with
\begin{equation} 
\label{cTk} 
\begin{split}
\hat{\cT}(k)  = \ii \left(\begin{array}{cc} \coth(k\delta) & 1/\sinh(k\delta) \\ -1/\sinh(k\delta) & -\coth(k\delta) \end{array}\right)
\end{split} 
\end{equation} 
and $\hat{\vu}(k) = (\hat u(k),\hat v(k))^t$ for $\vu(x)=(u(x),v(x))^t$.
Using this, it is easy to check that $\hat{\cT}(k)^2=-I$, which is equivalent to $\cT^2=-I$.

\begin{proof}[Derivation of \eqref{J}]
Suppose $0<a<2\delta$ and define the function $h(x)$ by
\begin{equation*} 
h(x)=\frac{\pi}{2\delta}\coth\left(\frac{\pi }{2\delta}(x -\ii a)\right).
\end{equation*} 
Even though $h(x)$ does not decay as $x \to \pm \infty$, the Fourier transform $\hat{h}$ of $h$ is well-defined as a tempered distribution. Indeed, the derivative 
\begin{equation*} 
h'(x) = -\bigg(\frac{\pi}{2 \delta \sinh(\frac{\pi  (x-\ii a)}{2 \delta})}\bigg)^2
\end{equation*} 
has exponential decay as $x \to \pm \infty$ and has a double pole at $x = \ii a + 2\ii \delta n$ for each integer $n$. Its Fourier transform  $\widehat{(h')}$ can be computed by a residue computation. 
The Fourier transform $\hat{h}$ can then be obtained for $k \neq 0$ by $\hat{h}(k) = \widehat{(h')}(k)/(\ii k)$. A similar computation applies if $-2\delta < a < 0$, and we arrive at \eqref{J}. 
\end{proof} 

\subsubsection{Eigenfunctions} 
Since $\cT^2=-I$, the eigenvalues of $\cT$ are $\pm \ii$. We now construct the corresponding eigenfunctions. 

By straightforward computations we obtain the following eigenvectors of the  matrix $\hat{\cT}(k)$ in \eqref{cTk}, 
\begin{equation} 
\hat g(k)\left(\begin{array}{c} \ee^{\pm k\delta/2} \\ - \ee^{\mp k\delta/2} \end{array}\right)
\end{equation} 
with corresponding eigenvalues $\pm\ii$, for an arbitrary function $\hat g(k)$ of $k$.
To get eigenfunctions of $\cT$ with appropriate analyticity properties, we restrict ourselves to functions $\hat g(k)$ such that $\hat g(k)\ee^{k \alpha}$ has a well-defined inverse Fourier transform $g(x-\ii \alpha)$ in a strip $-A < \alpha <A$ with $A>\delta/2$. 
For such functions,
\begin{equation}
\int_{\R}\frac{\dd{k}}{2\pi} \hat g(k)\ee^{\pm k\delta/2} \ee^{\ii kx} = g(x\mp \ii\delta/2),
\end{equation}
and the eigenfunctions of the operator $\cT$ are therefore as follows: {\em For arbitrary complex valued functions $g(z)$ of $z\in \C$ analytic in a strip $-A<\Im(z)<A$ with $A>\delta/2$, the vector valued functions
\begin{equation} 
\label{vpm0}
\vv_\pm(x) \equiv \left(\begin{array}{c} g(x\mp \ii\delta/2)  \\ -g(x\pm \ii\delta/2) \end{array}\right)
\end{equation} 
satisfy 
\begin{equation} 
\label{vpm1}
\cT\vv_\pm(x) =\pm\ii \vv_\pm(x).
\end{equation}}

\subsubsection{Pole ansatz}
\label{app:polh}
Inspired by the CMS-related soliton solutions known for the BO equation,\cite{CLP79,SAX08} we make the following ansatz to solve \eqref{2ILW2}, 
\begin{multline} 
\label{ansatz} 
\vu(x,t) =  \ii\sum_{j=1}^N\left(\begin{array}{c} \alpha(x-z_j(t)-\ii\delta/2)\\  -\alpha(x-z_j(t)+\ii\delta/2)   \end{array}\right) \\
  -\ii\sum_{j=1}^M\left(\begin{array}{c} \alpha(x-w_j(t)+\ii\delta/2)\\  -\alpha(x-w_j(t)-\ii\delta/2) \end{array}\right)
\end{multline}  
where $\alpha(x)=(\pi/2\delta)\coth(\pi x/2\delta)$, $N,M$ are arbitrary integers $\geq 0$, and with poles $z_j(t)$ and $w_j(t)$ to be determined. 
We note that, to obtain real-valued solutions, one must restrict this ansatz  to \eqref{solitonN}, i.e., $M=N$ and $w_j(t)=\bar{z}_j(t)$ for all $j$, but we find it convenient to derive a more general result. 
In the following, we sometimes write $z_j$ as shorthand for $z_j(t)$, etc.

The function $\alpha(z)$ is meromorphic with poles at $z=2\ii \delta n$, $n$ integer. 
Thus, if we restrict the imaginary parts of $z_j$ and $w_j$ as follows, 
\begin{equation} 
\label{cond} 
\Im(z_j\pm \ii\delta/2)\neq 2\delta n,\quad \Im(w_j\pm \ii\delta/2)\neq 2\delta n
\end{equation} 
for all integers $n$, then the result in \eqref{vpm0}--\eqref{vpm1} implies 
\begin{multline} 
\label{key} 
\cT\vu_{xx} =  
-\sum_{j=1}^N\left(\begin{array}{c} \alpha''(x-z_j-\ii\delta/2)\\  -\alpha''(x-z_j+\ii\delta/2)   \end{array}\right)\\
  -\sum_{j=1}^M\left(\begin{array}{c} \alpha''(x-w_j+\ii\delta/2)\\  -\alpha''(x-w_j-\ii\delta/2) \end{array}\right)
\end{multline} 
with $\alpha'(z)\equiv\partial_z\alpha(z)$ etc. 
We now use $\alpha(-z)=-\alpha(z)$ and the well-known identities\cite{C75} 
\begin{equation} 
\label{Idcoth} 
\begin{split} 
\alpha'(z)=-V(z),\quad 
\partial_z\big[\alpha(z)^2\big] = V'(z),\\
\alpha(z+2\ii\delta)=\alpha(z), \quad 
\partial_z\big[\alpha(z-a)\alpha(z-b)\big] \\ = \partial_z\big[\alpha(z-a)-\alpha(z-b)\big] \alpha(a-b), 
\end{split} 
\end{equation} 
with $V$ in \eqref{V}, and for arbitrary $z,a,b\in\C$. Using this we compute  
\begin{multline*} 
\vu_t +  (\vu.\vu)_x + \cT \vu_{xx} =  \sum_{j=1}^N\left(\begin{array}{c} V(x-z_j-\ii\delta/2)\\  -V(x-z_j+\ii\delta/2)   \end{array}\right)\\
\times \left( \ii \dot z_j +2\sum_{k\neq j}^N\alpha(z_j-z_k) - 2\sum_{k=1}^M \alpha(z_j-w_k + \ii\delta) \right) \\
  +\sum_{j=1}^M\left(\begin{array}{c} V(x-w_j+\ii\delta/2)\\  -V(x-w_j-\ii\delta/2) \end{array}\right)\\
\times \left( - \ii \dot w_j + 2\sum_{k\neq j}^M\alpha(w_j-w_k) - 2\sum_{k=1}^N \alpha(w_j-z_k+\ii\delta) \right)
\end{multline*} 
(the computations leading to this result are nearly the same as in the BO case\cite{SAX08} and thus omitted). 
This implies the following result: {\em The function in \eqref{ansatz} satisfies the nonchiral ILW equation in \eqref{2ILW2} provided 
the following system of equations is satisfied, 
\begin{equation} 
\label{BT} 
\begin{split} 
\dot z_j &= 2\ii\sum_{k\neq j}^N\alpha(z_j-z_k) - 2\ii \sum_{k=1}^M \alpha(z_j-w_k + \ii\delta),\\
\dot w_j &= -2\ii \sum_{k\neq j}^M\alpha(w_j-w_k) + 2\ii \sum_{k=1}^N \alpha(w_j-z_k+\ii\delta) , 
\end{split} 
\end{equation} 
and the conditions in \eqref{cond} hold true.}
 
The system in \eqref{BT} is known as a {\em B\"acklund transformation} for the hyperbolic CMS system.\cite{W82} 
It implies two decoupled systems of Newton's equations, 
\begin{subequations} 
\label{Newton1} 
\begin{align} 
\ddot z_j &= -\sum_{k\neq j}^N 4V'(z_j-z_k)\quad (j=1,\ldots,N), \label{Newton1A}\\
\ddot w_j & = -\sum_{k\neq j}^M 4V'(w_j-w_k) \quad (j=1,\ldots,M) \label{Newton1B}
\end{align} 
\end{subequations} 
with $V$ as in \eqref{V}; see Ref.~[\onlinecite{KP17}] for a recent alternative derivation of this result.
We thus obtain the following generalization of the result stated in the main text: {\em For arbitrary nonnegative integers $N,M$ and complex parameters  $a_j$, $j=1,\ldots,N$, and $b_j$, $j=1,\ldots,M$,  satisfying  
\begin{equation} 
\Im(a_j\pm \ii\delta/2)\neq 2\delta n,\quad \Im(b_j\pm \ii\delta/2)\neq 2\delta n
\end{equation} 
for all integers $n$, the function $\vu(x,t)$ in \eqref{ansatz} is a solution of the nonchiral ILW equation \eqref{2ILW2} provided the poles $z_j(t)$ and $w_j(t)$ satisfy Newton's equations for the hyperbolic CMS model in \eqref{Newton1} with initial conditions
\begin{equation*} 
\begin{split} 
&z_j(0)=a_j,\quad w_j(0)=b_j, \\
 \dot z_j(0) &=2\ii \sum_{k\neq j}^N\alpha(a_j-a_k) - 2\ii \sum_{k=1}^M \alpha(a_j-b_k+\ii\delta),\\
  \dot w_j(0) &=-2\ii\sum_{k\neq j}^M\alpha(b_j-b_k)+2\ii\sum_{k=1}^N \alpha(b_j-a_k+\ii\delta) . 
\end{split} 
\end{equation*} 
}
Restricting to $M=N$ and $b_j=\bar{a}_j$ for all $j$, we obtain the result stated in the main text (note that, in this special case, the initial conditions imply $w_j(t)= \bar z_j(t)$ for all $t$). 

A technical remark is in order. Strictly speaking, we proved the result above only for times, $t$, where the conditions in \eqref{cond} hold true. 
We did not point out this restriction before since we believe that, if the conditions in \eqref{cond} and \eqref{BT} hold true at time $t=0$, then the solutions $z_j(t)$ and $w_j(t)$ of \eqref{Newton1} satisfy the conditions in \eqref{cond} for all $t>0$. 
We checked this in several special cases by integrating \eqref{Newton1} numerically. 
We expect that this can be proved in general using the known explicit solution of the hyperbolic CMS model obtained with the projection method;\cite{OP81}
this is left for future work.

\subsection{Elliptic case}
\label{app:ell} 
We give details on how the derivation in Appendix~\ref{app:hyp} generalizes to the $L$-periodic case. 

\subsubsection{Periodic nonchiral ILW equation} 
To see that \eqref{2ILW} with $T$, $\tilde{T}$ in \eqref{TTe}--\eqref{MLe}  is the correct $L$-periodic generalization of the nonchiral ILW equation, one can check that \eqref{hatTT} still holds true but with Fourier modes, $k$,  restricted to integer multiples of $(2\pi/L)$, and for $L$-periodic functions $f(x)$ that have zero mean, $\hat f(0)\equiv \int_{-L/2}^{L/2} f(x)\dd{x} = 0$. 
Thus, $\cT^2=-I$, and the result in \eqref{vpm0}--\eqref{vpm1} holds true as it stands provided the function $f(z)$ is $L$-periodic, has zero mean, and is analytic in a strip $-A<\Im(z)<A$ for $A>\delta/2$. 
In particular,
\begin{multline} 
\label{vpm3} 
\cT \partial_x^2 \left(\begin{array}{c} \zeta_2(x-z\mp \ii\delta/2)  \\ -\zeta_2(x-z\pm \ii\delta/2) \end{array}\right) \\
= \mp \ii \left(\begin{array}{c} \wp'(x-z\mp \ii\delta/2) \\ -\wp'(x-z\pm \ii\delta/2) \end{array}\right)
\end{multline} 
using $\zeta''_2(z)=-\wp'(z)$. 
We can use this to construct soliton solutions related to the elliptic CMS model defined by Newton's equations \eqref{Newton} with the potential 
\begin{equation} 
\label{Vell} 
V(x) = \wp(x) . 
\end{equation}

\subsubsection{Pole ansatz}
The discussion above suggests to use the pole ansatz in \eqref{ansatz} with $\alpha(x)$ equal to $\zeta_1(x)$. 
However, this choice does not work since the third identity in \eqref{Idcoth} is not satisfied. The choice that works is   
\begin{equation} 
\alpha(x) = \zeta_2(x)
\end{equation} 
since $\zeta_2(z)$ is $2\ii\delta$-periodic. However, $\zeta_2(z)$ is not $L$-periodic: $\zeta_2(z+L) = \zeta_2(z)+c$ for some nonzero constant $c$. Thus, $\vu(x+L,t)=\vu(x,t)+\ii(N-M)(c,-c)^t$,  and,  to get a $L$-periodic function $\vu(x,t)$,  we must restrict to $M=N$. 

We use \eqref{vpm3} to obtain  
\begin{multline} 
\label{keyell} 
\cT\vu_{xx} =   \sum_{j=1}^N\left(\begin{array}{c} V'(x-z_j-\ii\delta/2)\\  -V'(x-z_j+\ii\delta/2)   \end{array}\right) \\
  + \sum_{j=1}^N\left(\begin{array}{c} V'(x-w_j+\ii\delta/2)\\  -V'(x-w_j-\ii\delta/2) \end{array}\right)
\end{multline} 
with $V$ in \eqref{Vell}.
We define $f_2(z)\equiv \partial_z[\zeta_2(z)^2-\wp(z)]$ and observe that the generalizations of the second and fourth identities in \eqref{Idcoth} are 
\begin{equation} 
\label{Id2} 
\partial_z\alpha(z)^2 = V'(z) + f_2(z)
\end{equation} 
and 
\begin{equation} 
\label{Id4} 
\begin{split} 
\partial_x\bigl[\alpha(x-a)\alpha(x-b)\bigr]
=\partial_x\big[\alpha(x-a) -\alpha(x-b)]\\ \times \alpha(a-b) +\frac12\bigl[f_2(x-a)+f_2(x-b)\big], 
\end{split} 
\end{equation} 
respectively (the latter follows from the following well-known functional equation satisfied by the Weierstrass functions,\cite{WW40}
$$[\zeta(x)+\zeta(y)+\zeta(z)]^2 = \wp(x)+\wp(y)+\wp(z)$$ provided $x+y+z=0$).
The first and third identities in \eqref{Idcoth} hold true as they stand. 

While $f_2(z)=0$ in the hyperbolic case, it is a nontrivial function in the elliptic case.
However, going through the computations described in Appendix~\ref{app:polh}, one finds that they generalize straightforwardly to the elliptic case provided $M=N$ (that \eqref{BT} for $M=N$ implies \eqref{Newton1} even in the elliptic case has been known for a long time\cite{W82}).
One thus obtains the same result as in the hyperbolic case but with the restriction $M=N$.

\section{Numerical method}
\label{app:B} 
We verified our soliton solutions numerically by adapting a method developed for solving the standard ILW equation\cite{PD00} to the nonchiral ILW equation \eqref{2ILW}. 
The numerical method applies to the periodic problem on the interval $[-L/2,L/2]$; for  initial conditions and for times, $t$, such that $u(x,t)$ and $v(x,t)$ are significantly different from zero only in an interval $[-\ell/2,\ell/2]$ with $0<\ell\ll L$,  this is an excellent approximation for the nonperiodic problem on $\R$. 
We thus checked numerically various 2- and 3-soliton solutions both for the periodic and nonperiodic problem, and we found excellent agreement. 
For example, the two-soliton solution in Fig.~\ref{Fig2} computed with our numerical method cannot be distinguished with bare eyes from the one obtained with our analytic result. 
We mention in passing that our numerical method is much more stable for initial conditions which give rise to soliton solutions than for generic initial conditions. 
In what follows, we describe our numeric method in more detail.

We employ the discrete Fourier transform
\begin{equation}
\begin{split} 
&u(x,t)\approx\sum\limits_{n=-N}^{N-1} \hat{u}_n(t) \ee^{\ii k_n x},\quad k_n\equiv n\frac{2\pi}{L} \\
&\hat{u}_n(t)=\frac{1}{2N}\sum_{j=-N}^{N-1}u(x_j)\ee^{-\ii k_n x_j},\quad x_j\equiv j\frac{L}{2N}
\end{split}
\end{equation} 
and the Fourier multiplier representations \eqref{hatTT} of the singular integral operators \eqref{TT}, 
\begin{equation} 
\label{hatTTn} 
\begin{split} 
&\widehat{(Tu)}_n(t) = \ii\coth(k_n\delta)\hat u_n(t),\\
&\widehat{(\tilde{T}u)}_n(t)  = \ii\frac1{\sinh(k_n\delta)}\hat u_n(t), 
\end{split} 
\end{equation} 
to obtain a system of ordinary differential equations for the time evolution of the Fourier coefficients via a semi-discrete collocation approximation\cite{PD00} 
(note that $\hat u_n(t)/L$ can be identified with the Fourier transform $\hat u(k_n,t)$). 
The numerical approximation for the nonlinear terms is $\widehat{(2u u_x)}_n(t)=\ii k_n \widehat{(u^2)}_n(t)$ with 
\begin{equation} 
\widehat{(u^2)}_n(t) \approx \sum_m \hat u_{n-m}(t)\hat u_m(t)\quad (-N\leq n\leq N-1)
\end{equation} 
where the sum on the right-hand side is over the integers $m$ in the range $-N\leq m\leq N-1$ such that $-N\leq n-m \leq N-1$. 
In our tests we used $L=200$ and $N=512$. 

\bibliographystyle{apsrev}

\begin{thebibliography}{99}
\bibitem{rem1} ``CMS systems'' is short for ``$A_{N-1}$-type CMS systems''.
\bibitem{remg} The factor 4 in Eq.\ \eqref{Newton} etc.\ is the CMS coupling constant, $g^2$ for $g=2$; our choice $g=2$ is for convenience, and it can be easily changed by rescaling time.
\bibitem{rem3} One usually adds a term $\frac1{\delta} u_x$ to the left-hand side of the ILW equation to ensure that it reduces to the Korteweg-de Vries equation in the limit $\delta\to 0$. 
This term is trivial in that it can be removed by a Galilean transformation, or by the transformation $u\mapsto u-1/2\delta$, and we find it convenient to ignore it.
\bibitem{remPDE} We interpret ``nonlinear PDEs'' here in a broader sense, including nonlinear integro-differential equations.
\bibitem{remILW} The parameter $\delta$ in the KdV dispersion relation can be scaled to any value, but we include it for dimensional reasons, and also to indicate the relation to the ILW dispersion relation: $k^2\coth(k\delta) = -k/\delta + k^3\delta/3+\ldots$ for small $\delta$, where the first term corresponds to the term $u_x/\delta$ that we ignore,\cite{rem3}, and the second is the KdV dispersion relation. Note that this argument does not prove that the ILW equation reduces to the KdV equation in the limit $\delta\to 0$: to prove this, one has to remove $\delta$ in the first term by adding $u_x/\delta$ and from the second term by scaling, and then take the limit $\delta\to 0$.
\end{thebibliography}

\end{document}